# Children's Overtrust and Shifting Perspectives of Generative AI

Jaemarie Solyst, Carnegie Mellon University, jsolyst@andrew.cmu.edu
Ellia Yang, Carnegie Mellon University, elliay@andrew.cmu.edu
Shixian Xie, Carnegie Mellon University, shixianx@andrew.cmu.edu
Jessica Hammer, Carnegie Mellon University, hammerj@andrew.cmu.edu
Amy Ogan, Carnegie Mellon University, aeo@andrew.cmu.edu
Motahhare Eslami, Carnegie Mellon University, meslami@andrew.cmu.edu

**Abstract:** The capabilities of generative AI (genAI) have dramatically increased in recent times, and there are opportunities for children to leverage new features for personal and school-related endeavors. However, while the future of genAI is taking form, there remain potentially harmful limitations, such as generation of outputs with misinformation and bias. We ran a workshop study focused on ChatGPT to explore middle school girls' (N = 26) attitudes and reasoning about how genAI works. We focused on girls who are often disproportionately impacted by algorithmic bias. We found that: (1) middle school girls were initially overtrusting of genAI, (2) deliberate exposure to the limitations and mistakes of generative AI shifted this overtrust to disillusionment about genAI capabilities, though they were still optimistic for future possibilities of genAI, and (3) their ideas about school policy were nuanced. This work informs how children think about genAI like ChatGPT and its integration in learning settings.

## Introduction

Many children today are exposed to artificial intelligence (AI) in their everyday lives, in the myriad of online platforms and apps that they regularly interact with. Generative AI (genAI), a type of AI which can create new content rather than predicting or classifying existing information, has led to increasing interest in AI-driven technologies. Tools such as ChatGPT can generate text-based outputs, while MidJourney and Dall-E can generate image outputs. The launch of ChatGPT has popularized genAI for everyday users, youth included. For instance, Google announced in November 2023 that they made their genAI fitting to be used by teens.

While hailed for its transformative potential, genAI also raises many concerns regarding its propensity for misinformation and demonstrating algorithmic bias. Misinformation can be defined as information or opinions not supported by experts or evidence (Vraga & Bode, 2020). Algorithmic bias can be thought of as systems unfairly discriminating against certain individuals or groups of people (Friedman & Nissenbaum, 1996). For example, AI-generated news has been found to contain fake facts (Wiggers, 2023). Additionally, generated images and text can contain bias. Qadri et al. (2023) found that text-to-image generative AI represented South Asia in ways that were harmful, stereotypical, and Western-centric. While research on adults shows skepticism toward genAI news content (Longoni et al., 2022), there is a timely need to build on sparse literature exploring children's perceptions of genAI, as they may be more susceptible to misinformation (Sutherland & Hayne, 2001). With many recent efforts toward AI literacy in K-12 (Casal-Otero et al., 2023), there is a need to understand how genAI may fit into their broader AI understandings and learning experiences. Ali et al. (2021) explored genAI with middle school students with a focus on generated text and visual content. They found that youth were interested in creative options that genAI could support but were concerned about how misinformation could be spread, e.g., through the use of DeepFakes. This study took place early in genAI development, in which it often produced nonsensical outputs. Given dramatic recent improvements in genAI, we aim to build on this in understanding youth perspectives. We use ChatGPT as our context, given its popularity at the time of this study.

We contextualize this research with prior work on children's understandings of conventional (non-genAI) conversational AI (e.g., used in chatbots or agents like Alexa and Siri). Prior literature suggests that young children (3-6 years old) do not view conversational AI as human or purely machine, rather something in between (Xu & Warschauer, 2020), but fewer studies have older youth in the age range of this study. Other work has found that with age (Flanagan et al., 2023) and exposure to programming AI (Druga & Ko, 2021), children can gain understanding of the capabilities and limits of smart devices. However, they often overestimate the intelligence capabilities of conversational AI or believe that it is smarter than them (Druga et al., 2018). Further, attribution of human traits to AI, such as having morals, sometimes hinders their ability to accurately understand its capabilities (Girouard-Hallam et al., 2021). In conjunction, prior work suggests that children may be vulnerable to overtrusting conversational AI. As genAI chatbots may be particularly convincing of having sentience, it is important to understand how children may (dis)trust new genAI technology, especially since it can produce misinformation in ways that are novel to users.

Children may also interact with genAI in schools, where teachers and administrators have shown both optimism and apprehension about genAI. While New York City and others have banned ChatGPT in schools altogether (Rosenblatt, 2023), some teachers are taking to the internet to collaborate on new ideas for assignments that incorporate genAI. Many adults believe that students should be able to use and leverage genAI for educational benefits (Crawford et al., 2023). Simultaneously, there is concern around students using the same technology in ways that could compromise academic integrity (Cotton et al., 2023). Given the likelihood of encountering genAI at school or interest in using genAI for homework, there is a need to better understand youth perspectives.

In this study, we worked with girls in middle school, as middle school is a key time to develop STEM identity (Sadler et al., 2012). Our focus was on girls, since girls and women are underrepresented in creating tech but are disproportionately impacted by AI bias, such as facing gender injustice facilitated by AI in biased search results (Yin & Sankin, 2020) and voice recognition working less for feminine voices (Bajorek, 2019). In terms of misinformation detection, prior literature has suggested that there are no significant differences between girls' and boys' abilities to recognize misinformation (e.g., Morais & Cruz, 2020).

We asked three research questions: How do middle school girls (1) perceive text-based generative AI, and how is this affected by results that show misinformation or bias? (2) reason about genAI's capabilities and learn about its limitations? (3) think school classroom policies should address genAI?

We found that the participants did not distinguish genAI from aspects of popular conventional AI (e.g., Google search) or other computing tools like calculators. Early on, participants seemed to overestimate ChatGPT's capabilities but were more discerning after our workshop. They were thoughtful about school policy and shared concerns that are similar to what adult teachers are currently grappling with. This work informs the literature on how girls think about genAI in order to understand what potential knowledge gaps need to be addressed. We offer recommendations for how to support children in building working mental models, including the limitations of genAI, and describe further implications for developing school policies.

## Methods

### Participants and recruitment

To collect data, we ran an IRB-approved educational workshop study with two different groups, "ScienceStars" and "CompuJam," with 26 participants in total. One-off workshops were integrated into established outreach programs aimed at STEM exposure for middle school girls. Our session lengths were determined by the existing program infrastructure. Both programs were held at a university in a mid-sized city on the East Coast of the United States. All students had heard of "artificial intelligence," five (2 in CompuJam, 3 in ScienceStars) had heard of "generative AI," and ten (6 in CompuJam, 4 in ScienceStars) had heard of "ChatGPT."

ScienceStars (N=14) was a free event with open enrollment for middle school girls, which took place all day on a Saturday. The organization delivering the program recruited participants by sending information about their event to local schools. The event included several technology-related lessons and activities that participants were assigned to by the program organizers. Our 45-minute workshop was one of the sessions. The decision was made with the research team, program organizers, and IRB not to compensate students for their participation due to the program structure. In this event, there were multiple STEM topic sessions run at the same time, and only some students were randomly assigned to our session during that time block.

CompuJam (N=12) was a free program which accepted all middle school girls who applied to participate in the after-school STEM program. Previous sessions of the weekly program included workshops on computing topics. Students who were a part of the program came to our 90-minute session and had prior exposure to technical topics. Recruiting was done by organizers who emailed school counselors and teachers and asked them to pass on program information to families. Again, students who participated in our session were not compensated, as program organizers and staff diversity coordinators did not see payment as aligned with their goals.

### Workshop content

Below we describe the four parts of the workshop: (1) an introduction to genAI with reflection on benefits, drawbacks, and algorithmic bias, (2) exposure to genAI limitations through a guessing game activity, (3) imagining future applications of genAI, and (4) discussion around fair policies pertaining to access and school.

*GenAI Introduction and Reflection on Bias.* After a brief introduction to AI and genAI through educational slides and demonstrating images generated by Dall-E from different prompts, we gave the students an introduction to the concept of ChatGPT and showed outputs. To understand initial reactions to this technology and then reasoning around genAI bias, we started an open-ended group discussion with all the students in the room about potential benefits and drawbacks of ChatGPT. We then introduced the notion that genAI could have bias by showing a response generated by ChatGPT in which it was asked to generate a gift list for girls and then boys. Without showing the prompt, the outputs were shown to the students, and they were asked to guess which list was for girls and which was for boys and then to provide comments on whether they thought this was biased.

*Guessing Game (limitations of genAI).* Next, we had a Guessing Game activity that was designed to show the participants that ChatGPT could also sometimes return incorrect outputs. We aimed to understand how exposure to the technology's limitations could impact their perspectives of genAI. Students were shown users' inputs and ChatGPT outputs that were sometimes incorrect. Input prompts and outputs were shown in the same order as listed in Table 1. We aimed to vary questions in terms of type of input and output (the "Why We Selected" column also explains the rationale), and all were related to facts (not opinions) for the scope of the activity.

For each question, the students were asked whether they thought the response from ChatGPT was correct, incorrect, or if they were unsure, and to state why. They all had their own game answer sheets to fill out individually. We then revealed the answers one by one and discussed as a group. After this, we asked again about the benefits and drawbacks of ChatGPT.

**Table 1**
*Guessing Game: User Inputs and ChatGPT Outputs*

| User Input | ChatGPT Output | Answer | Why We Selected |
|---|---|---|---|
| (Q1) "Does [study city] have the most bridges in the world?" | There are other cities with more bridges. | Correct | Grounded example in learners' context |
| (Q2) "Compute 32874*34918" | 1147010 (shows multiplication process) | Incorrect | Common example from news |
| (Q3) "List related papers on machine learning" | A list of seven papers with title, author, year of publication | Incorrect | Common example from news |
| (Q4) "How do I add a line break to a comment in Google docs?" | A four-step process on what keys to press in what order. | Incorrect | Exploratory procedural question |
| (Q5) "When did Ohio fight Pennsylvania?" | Ohio and Pennsylvania have never fought a war. | Correct | Incorrect or ambiguous input |

**Figure 1**
*Screenshots of Incorrect ChatGPT Outputs for Q2 and Q4 in the Guessing Game Activity*

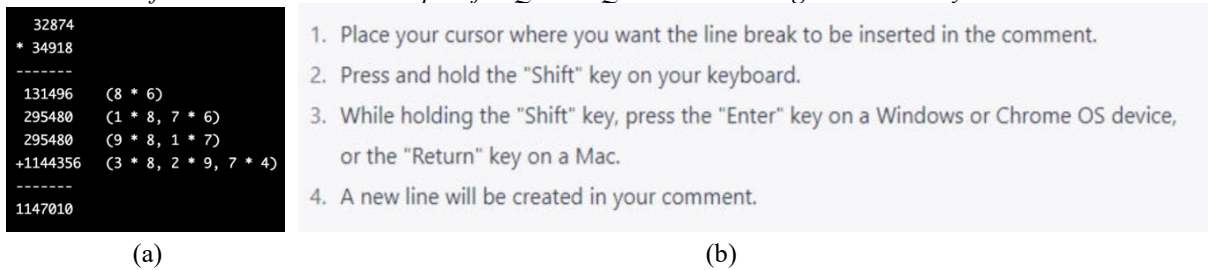

(a)                                                                 (b)

*Imagining Future Applications with GenAI.* The third activity, GenAI Futures, was a printed final project booklet in which the participants ideated future applications of genAI for a problem that they wanted to solve. Participants could write or draw their answers to the prompts: "I would like to use generative AI to solve this problem," "The types of generative AI I would use are," and "Prompts I would give generative AI in order to solve my problem." We discussed the students' work and ideas by having open-ended one-on-one conversations with each participant about their booklets. This activity was only run with CompuJam due to having more time.

*Fair Policies.* In both sessions, we held an open-ended group discussion on school policies and payment policies for genAI. To address genAI in school, students were prompted with a scenario in which their classmate was using ChatGPT on an essay for a homework assignment and asked whether this was fair. They were then told about how some school districts have already banned the technology and asked their opinions on banning ChatGPT in schools. We also introduced ChatGPT Plus, which was the paid tier of ChatGPT where users could always have access to the chatbot. We held an open-ended discussion on whether this was fair or not.

## Data capture and analysis

Due to program and participant preferences, we did not record the entire session. We had three or four researchers support the data collection; one acted as the main session leader and the others helped to take detailed notes and mark down observations, including what students said and their sentiments. Researchers additionally video and audio recorded one-on-one conversations about students' booklets if they consented to it.

We conducted consensus-based (Hammer & Berland, 2014) qualitative thematic analysis (Braun & Clarke, 2012) with our team on the data sources: students' artifacts, one-on-one recordings, and researcher notes of the sessions. The first and second author lead the data analysis and regularly checked in with the team to make iterations. Discussions and discrepancies were held to finalize the results.

*Limitations.* There are two main limitations in this study design. First, due to program regulations and preferences, we could not gather in-depth participant demographic backgrounds although both groups showed an appearance of racial diversity. Given that participants self-selected into the event and their parents could transport them to the university, they may have come from higher-resourced backgrounds and have higher exposure to STEM topics. Additionally, we only were allowed to show static images of ChatGPT. Importantly, a discussion with adult stakeholders in both programs raised concerns about using ChatGPT live, so all ChatGPT outputs shown throughout the workshops were screenshots. Given that the materials were real ChatGPT outputs, we believe that this allowed us to credibly capture the participants' beliefs, however it is important to contextualize their responses as reacting to engineered prompts rather than more conversational interactions.

*Positionality.* We recognize that our positionality and identities interacted with the analyses. We come from a range of academic, racial, cultural, and socioeconomic backgrounds. Our academic backgrounds include computing, human-computer interaction, learning science, and culturally responsive computing. These impacted the lenses that authors brought to analyses and the final set of agreed upon themes. All authors on this paper are women, which impacted how we could relate to the girls as the study population.

## Findings

### Children's mental models: Blurring the lines between genAI, conventional AI, and computational tools

When students were prompted to describe their understanding of ChatGPT as an explanation for how it answered the questions in the guessing game, the most common models referenced AI assistants, "*like an advanced Alexa,*" or a search engine, "*similar to Google or other [search engines], which are often right these days.*" These ideas also reappeared when they were prompted to think about the payment wall for ChatGPT, with some students weighing if it would make sense to pay for it if it was so similar to Google search — "*What makes it better than Google if you have to pay for it?*"

The students correctly believed that genAI was limited by what data it had access to, just as with search engines and other conventional systems. When reasoning about guessing game Q3 (science resources), a student pondered, "*Depending on when you ask the question, the information making it true or false might not be in the software at the time. Like ... we don't know when the research was published and if there's a new counterargument.*" In other words, if the information was not available online, then genAI could not be expected to use it; the same would be true for a search engine.

On the other hand, they believed that genAI should be good at fact-based or computational tasks, since it was a computing technology. The learners were particularly surprised when the genAI output was incorrect for the multiplication problem, since they thought it should at least be capable of what a calculator could do. One learner commented, "*It probably is [correct], but it is a very big number for me to be sure. Anyways, calculators can easily get the answers and have for many years.*" After the answer was revealed, another student found it "*very shocking, [because she] thought there was no chance it would get a math question wrong.*" This incorporation of elements of genAI, conventional AI, and non-AI computational tools suggests that the participants held a *pastiche model* (Collins and Gentner, 1987) — that is, a mental model composed of two or more potentially inconsistent representations of how a system operates.

### Overtrust influenced by aesthetic legitimacy and perceived transparency

In the guessing game, we observed a notable trend of overtrust in ChatGPT's responses. Many students either deemed the incorrect responses of ChatGPT as correct, or they expressed uncertainty. Despite being previously informed about ChatGPT's potential for generating biased outputs in the introduction and reflection activity, the learners showed a tendency to accept the AI's answers at face value. Specifically, for questions Q2, Q3, and Q4, where ChatGPT generated incorrect answers, a notable 59% of the students' assessments inaccurately identified these responses as correct (Table 2 and Figure 2 show details about their assessments for these questions in both workshop groups). In their explanations of their assessments, learners relied on superficial information about the outputs to reason about correctness. We saw two main factors that led to falsely perceiving incorrect outputs as correct: a) visual appearance with additional seemingly correct information, which we call *aesthetic legitimacy*, and b) *perceived transparency*, the inclusion of information about the AI's reasoning process, albeit misleading.

*Aesthetic legitimacy*. First, learners were persuaded by the visual appearance and layout of information which gave the incorrect ChatGPT outputs legitimacy. This was especially the case for guessing game Q4 (tech help), where they suggested that it was correct because "*it labels out the steps*" with numbers (Figure 1). Given that the question was about a procedure, this format was particularly convincing for learners. In addition, when other seemingly correct information was presented beyond the answer, even when erroneous, it led them to believe the output was overall correct. For example, one student reasoned that ChatGPT had "*good facts behind [the answer, so] it's probably right,*" Another commented that the output was correct, since "*the answer shows dates and specific years,*" and "*the titles relate.*"

**Table 2**
*Percentage of Learners Who Perceived the Incorrect ChatGPT Responses as Correct*

| Group | Q2 (Multiplication) | Q3 (Science Resources) | Q4 (Tech Help) | Combined |
| --- | --- | --- | --- | --- |
| Science Stars | 36% | 58% | 79% | 57% |
| CompuJam | 67% | 50% | 67% | 61% |
| Total | 51% | 55% | 73% | **59%** |

**Figure 2**
*Graphs of How Many Learners Answered Incorrect, Unsure, or Correct in the Guessing Game by Question*

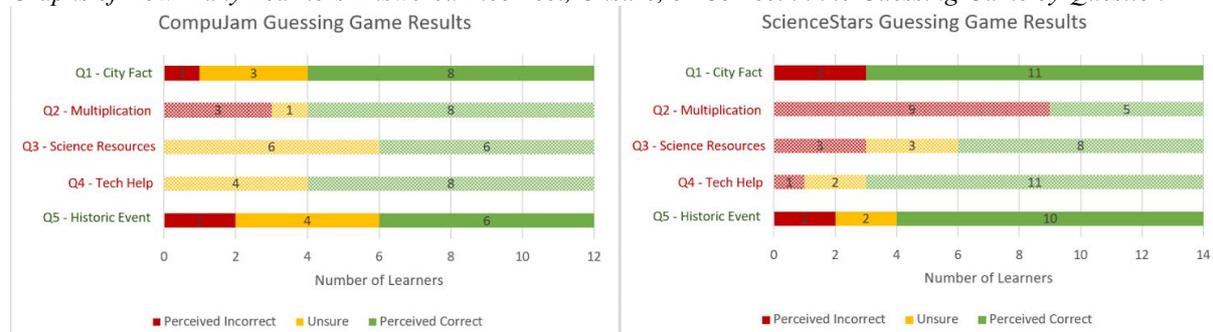

*Perceived transparency*. Learners were also influenced by what they perceived as transparency. They suggested that when the genAI output looked like it showed its reasoning, such as in Q2 (long multiplication problem), it was correct. For example, one student suggested that "*it did the process of normal multiplication.*" We note that this transparency is only a perception, as ChatGPT may have provided reasonings in outputs completely independent of its final responses.

However, perceived transparency also had the potential to trigger the opposite effect. In ScienceStars but not CompuJam (Figure 2), some learners noticed issues with the way ChatGPT solved the multiplication problem, which led them to check the math problem themselves (Figure 1). Others were then influenced by observing this process — once they saw peers trying to check the multiplication, they also tried it out themselves. This allowed them to fact-check ChatGPT using their own prior knowledge. As a result, a majority of the participants correctly identified ChatGPT's answer as being wrong. In contrast, very few in CompuJam attempted the multiplication themselves; many of these students incorrectly assumed that ChatGPT's answer to the multiplication problem was accurate based on their model of ChatGPT as a computational tool. Therefore, we saw that transparency had two seemingly opposite impacts for different learners. Some found its mere presence convincing, while for learners who were skeptical or observed others checking the answer, it helped them respond critically.

### Shift from overtrust to understanding genAI limits and future possibilities

After we discussed the incorrect ChatGPT answers as a group, and students became more aware of the drawbacks of the technology, their perspective of ChatGPT's capabilities shifted: they started questioning genAI. This shift was evident in the pre- and post-brainstorming discussion about benefits and risks of genAI like ChatGPT, as well as when the learners considered whether they would pay for genAI services. When discussing benefits and risks at the start of the session, learners had no trouble thinking of exciting beneficial applications, like creativity support, solving problems, and homework help. However, they struggled to think of drawbacks. With further prompting, they reported high-level issues like concern about overuse of genAI — "*Humans could become too dependent, and you wouldn't use your own brain*." After the guessing game, students could more quickly reflect on drawbacks, such as not using genAI in certain high-risk scenarios. One mentioned, "*If you use it for medical instructions, and it gets it wrong, then the person is dead.*" Another asked, "*But should we trust it with medical* *questions? If it can't even do a multiplication problem, why should we trust it with medical questions?*" This illustrates how an erroneous example can have large impacts on mental models and understanding limitations.

*Role of human-generated prompts*. Learners were aware that genAI outputs depend on the correctness and specificity of human input. This was brought to attention by guessing game Q5 (historic event), with a student commenting that the correctness could "*depend on if the [human user] input is good.*" Another learner mentioned that "*different people interpret [terms differently],*" so the output may not be aligned with what the user input intended. The role of human input was particularly salient when learners discussed how ChatGPT could have gender bias in the gift list prompt, 7 (2 CompuJam, 5 ScienceStars) brought up how it was "*stereotypical*" to gender norms and suggested that this type of bias was unethical. One learner suggested that the AI was limiting people in how they could represent themselves, and others suggested that it was generalizing interests for boys versus girls, since "*not everyone is going to be interested in the same things just based on gender*" (CompuJam). Students from both groups expressed that they did like some of the suggestions that the AI made for girls, but they also liked some of the gifts that the AI delegated only for boys, like sports gear, and felt that their interests should be considered more than their gender. However, another line of thought suggested that perhaps the onus was on the user for searching based on gender — "*[The list] may be biased, but if we wanted a more equal list, we shouldn't have specified the gender.*" We elaborate on work by Solyst et al. (2023), suggesting that youth may believe there is responsibility of the users to use the right input terms to work around algorithmic bias. This finding adds that learners believe bias in generated outputs can be reduced by not specifying specific identities.

*Embracing potential future possibilities*. Despite their perspective shift in genAI capabilities, learners remained hopeful about the future applications even after being made aware of and discussing its limitations. Directly after discussing drawbacks, students engaged in the future genAI booklet activity, where they suggested many high-stakes challenges that they believed genAI could address in the future, such as "*world hunger*" and "*therapy.*" We found this contrast striking. For example, although they had just raised their concerns about using genAI in medical scenarios, they still suggested it could be used for "*solving cancer.*" This suggests that awareness of limitations did not discourage from considering the hopeful possibilities of future genAI applications.

## Fair genAI policies in school

Learners understood that using genAI did not just impact individuals but communities and institutions. We saw this in the context of schools, since the girls envisioned one main benefit of genAI being homework help. They were interested in how genAI could impact schoolwork in both positive and negative ways. In terms of positive aspects, students brought up how genAI could fill gaps in learning, since it could *"help people understand things better"* or even be used *"as a teacher for homework."* They also saw how it could support lower resourced schools to offer "*better education and resources [to avoid] overspending*," and "*people who don't have the resources to go to school … can ask the AI questions.*" However, since the learners had just discussed misinformation, they were concerned about the implications on their schoolwork, mentioning that "*it could affect grade[s] negatively.*"

Similar to adult teachers' concerns, cheating with genAI in school was a main topic of discussion as well. Six of the twelve learners in CompuJam agreed with some cities' choices to ban ChatGPT. One student expressed concern that "*a lot of people could use ChatGPT to cheat, so blocking it could prevent cheating, like it could just write their essays or do their math questions for them.*" Some learners also commented on the impacts of using ChatGPT to cheat and suggested that this could result in students "*not pay[ing] attention as closely*" to assignments or "*relying on it*" too much. Other students differentiated between types of use, such as one who suggested that it "*depends on what you're using it for.*" Another student added that it "*would be unfair if [ChatGPT] writes the essay for you, but if it just helps, it's fair.*" Comparisons were made between other AI-driven technology already used in school settings, such as Grammarly — "*Isn't Grammarly an AI that helps you write essays? People already use it, so I think [using ChatGPT in school] would be fair.*"

Lastly, when discussing the payment wall for ChatGPT, some were concerned about this as a barrier for students. One learner reasoned that "*if only some people have to pay, then it would be unfair, but if it was all free, then it becomes fair.*" This is in line with work finding that youth were sensitive to how access to AI may exacerbate societal inequities and the idea of equality as fairness (Solyst et al., 2022; Solyst et al., 2023). However, another learner countered by suggesting that "*it's like capitalism. You have to pay for stuff you want, and it's not vital. It's like a service that makes life comfier.*" A few other learners agreed, suggesting that some believed that genAI may not be *"vital"* to the future of learning or the classroom.

## Discussion

This study explored girls' perspectives on genAI. We found that their conceptualizations of genAI integrated concepts from conventional AI (e.g., search engines and voice assistants) and computing tools like calculators. When evaluating ChatGPT outputs, students leaned toward trusting the information, except for when they were able and thought to check the answer themselves. We observed that learners used superficial characteristics to determine if the output was correct, including aesthetic legitimacy (the visual layout of information), the addition of other seemingly correct information, and perceived transparency. This transparency had two different effects, which led them to either trust the output more or be able to critically reason about the output to determine its correctness. After viewing how ChatGPT could make unexpected errors, learners had a better understanding of the limits of genAI. They also shared concerns regarding educational environments where genAI like ChatGPT could be misused or overused. Despite this, they were still optimistic about future applications of genAI.

*Generative AI disillusionment through exposure to AI limitations.* As our study progressed, we observed a shift in the students' trust of genAI. Initially imbued with an overtrust in its capabilities, their attitudes transformed upon being presented with examples of the genAI's errors and limitations. This corroborates previous research findings in adult populations, in which the exposure to algorithmic mistakes can help adults become disillusioned about the abilities of algorithmic systems. Disillusionment refers to realizing the limitations for a more realistic mental model of the algorithm at hand (Eslami et al., 2018). This also adds to prior literature suggesting that youth gained awareness of AI limits through the act of programming (Druga & Ko 2021); our findings suggest that skepticism can also arise simply from exposure to erroneous outputs. This exposure played a crucial role in tempering learners' overconfidence, fostering a more informed and critical viewpoint. It highlighted the importance of experiential learning in shaping children's understanding of technology's fallibility and the value of skepticism in the digital age. To further bolster critical thinking about genAI, there is a need to explicitly educate about how genAI is different from other conventional AI. Lack of differentiation between the two led the students to make false assumptions about genAI capabilities.

*Updated media literacy skills and mental models*. We found that the students were using an inaccurate mental model for understanding genAI, and they may need new media literacy skills to evaluate genAI content. For example, many current media literacy skills (which are not targeted towards genAI) suggest checking where sources are posted and determining their authors (Bulger & Davidson, 2018). However, these are not applicable to genAI output, since it is generated by the technology, and sources are obfuscated. This leaves youth vulnerable to misinformation from genAI. One possible approach to developing a genAI literacy that we have found is in our extension of the *erroneous example* paradigm from technology-enhanced learning (Tsovaltzi et al., 2010); that is, engaging learners in an example task that deliberately contains an error in order to support deeper reflection and in this case, conceptual change. With media literacy opportunities offered in schools, educators also may need to update their understandings and curriculums to cover genAI.

In conclusion, this study aimed to illuminate middle school girls' perspectives and knowledge gaps around genAI. We add onto prior literature on children's mental models of AI (e.g., Flanagan et al., 2023; Xu & Warshauer, 2020) in the context of genAI and the middle school age range for girls. Since girls and boys have similar abilities to recognize misinformation (Morais & Cruz, 2020), our findings on overtrust and reasoning about correctness of genAI outputs may be transferrable to a broader population. The guessing game activity used in this study could be employed in teaching AI literacy about genAI, since we saw how more accurate mental models may be developed through exposure to surprising erroneous examples. We saw learners started from overtrusting genAI but finished our workshop with the ability to ask critical questions about genAI's shortcomings, as well as consider when it should and should not be used.

## Acknowledgments

We thank the organizations and youth participants that made this study possible, as well as Yuki Chen who helped with the study. This work is supported by the Jacobs Foundation's CERES Network and the NSF DRL-1811086.